# Do All Asians Look the Same?: A Comparative Analysis of the East Asian Facial Color Desires using Instagram

Jaeyoun You

Intelligence and information, Seoul National University, Seoul, South Korea, you.jae@snu.ac.kr

Sojeong Park

Communication, Seoul National University, Seoul, South Korea, psj25psj25@snu.ac.kr

Seok-Kyeong Hong[*]

Communication, Seoul National University, Seoul, South Korea, skhong63@snu.ac.kr

Bongwon Suh[*]

Intelligence and information, Seoul National University, Seoul, South Korea, bongwon@snu.ac.kr

Selfies represent people's desires, and social media platforms like Instagram have been flooded with them. This study uses selfie data to examine how peoples' desires for ideal facial representations vary by region, particularly in East Asia. Through the analysis, we aim to refute the "all Asians prefer identical visuals," which is a subset of the prevalent Western belief that "all Asians look the same." Our findings, reinforced by postcolonial interpretations, dispute those assumptions. We propose a strategy for resolving the mismatch between real-world desires and the Western beauty market's views. We expect the disparity between hegemonic color schemes and the augmented skin colors shown by our results may facilitate the study of color and Asian identity.

**ACM Reference Format:**

First Author's Name, Initials, and Last Name, Second Author's Name, Initials, and Last Name, and Third Author's Name, Initials, and Last Name. 2018. The Title of the Paper: ACM Conference Proceedings Manuscript Submission Template: This is the subtitle of the paper, this document both explains and embodies the submission format for authors using Word. In Woodstock '18: ACM Symposium on Neural Gaze Detection, June 03–05, 2018, Woodstock, NY. ACM, New York, NY, USA, 10 pages. NOTE: This block will be automatically generated when manuscripts are processed after acceptance.

---

[*] Corresponding authors

# 1   Introduction

The selfie is an illustrative material and a medium for self-representation. With massive advances in smartphones, capturing a selfie has become a frequent behavior; thus, researchers have focused on individuals photographing (and photoshopping) their faces. Considerable psychological research has examined self-esteem and pathological narcissism with respect to selfie behaviors [6][51]. The human-computer interaction studies have reported editing behaviors for selfie takers' ideal self-expressions, dietary matters emerging from "visual diets," and body dysmorphic disorders [10, 45, 41, 5]. In addition, a philosophical research conducted a phenomenological analysis of the selfie as an embodied entity [56]. That is, selfies can be addressed as a source of desires.

Notably, the selfie applications have evolved over time to enable users to reflect more desires on their portraits. Apps such as "Snow" or "Foodie" have been performing facial color and attributes corrections. Furthermore, built-in camera apps on smartphones also include automatic modifications in the default mode. Although the detailed guidelines are unknown, some claim that these alterations follow explicit preferences of beauty [58][14]. Specifically, the preferred style of beauty is embedded and augmented by digital devices that mediate human representation in a virtual space.

In the subprocess of choosing the modification tools in apps, we presume that the hegemonic beauty standards are imposed on more people by recommending filter services. The effect of filters on the users has become stronger with gradual interactions. For example, some teenagers seriously asserted that they need virtual beauty filters to mask their real faces [37]. Even if we omit the extreme cases, we cannot avoid the underlying concept of the *augmented beauty* in selfies, meaning that virtual look – intentionally modified – distinct from the real-world appearance.

Because users' desires are conveyed in social spaces through augmented beauty, we may trace the origins of the "ideal" beauty system. Considering the notion of "white", in which there exists an ideological preference for the hegemony of the Western seen-as-white expressed by skin color, it is likely that the filtered selfie color also demonstrates specific patterns [16]. However, do non-Westerners' selfies really show the whiteness? Even if that is true, does this mean that everyone expresses themselves the same way, regardless of their cultural backgrounds?

This study investigated how various skin color patterns reflecting desires are revealed through an examination of selfies taken in each East Asian country. Our strong hypothesis was that the appearance of all Asians is not the same, contrary to the conventional wisdom of the Western culture. Additionally, we sought to establish that their desires for beauty are distinct. For examining, we analyzed three skin makeup attributes of faces in photos: tone, brightness, and texture. These parameters were deployed by the "Mibaek" makeup, a Korean way of skin-whitening that is currently leading beauty trend in East Asia; it emphasizes the flawless, transparent, and bright appearance of the skin [36].

Our primary research questions are as follows: First, does the appearance of the East Asian individuals' skin color in selfies vary depending on their nationality? Second, if there is a difference, how can it be distinguished and are there any patterns among countries regarding the same? Third, how are their representational color systems placed in dominating systems? What limitations can be identified in the current color scheme and can our findings contribute to them?

In the following section, we aimed to follow the postcolonial approach to skin color measurements and social media analysis. Furthermore, we explored Western stereotypes about East Asians that reduce them to being "yellow." Subsequently, we assessed the previous research on skin measurements that resulted in hegemonic color spaces, affected by the Westerners' perceptions. By outlining how people reveal their identities in social network spaces, we investigated whether their products (selfies) complement the limitations of the existing skin measurement research.

# 2   Postcolonial Approach to Social media Analysis

Prior to reviewing the skin color studies, we first address the concept of the postcolonial approach. Postcolonial theory and criticism mainly discuss dominating powers in multiple fields, including culture, politics, and economics. This theory examines the hegemonic effects in highly stratified social structures pays attention to socially marginalized groups. The postcolonial discourse should investigate the process of social systems that empowers the invisible upper class' style and reveal how the social minorities, or subalterns are deprived of their voices [47][31]. We refer to this approach to explore the hegemonic process that builds a dominating beauty system and shed light on Asian desires.

## 2.1 How East Asians Became Yellow

Skin color is inextricably linked to identity. White, as a complexion, does not merely refer to a physically recognized color or the equivalent RGB (red, green, blue) value of (255, 255, 255). Several attempts have been made in Whiteness studies and critical racial studies to examine white as a racial color and White people as a social construct [1][25][16][17]. As noted by Dyer, the Western media has constructed White individuals as "unraced," thereby consolidating and reproducing "white" as a human norm. The significance is that "whiteness is more an ascription than a fixed given" [16]. Those who have access to privileges have been deemed Whites, irrespective of how they truly appear, whereas those who do not, have been regarded as colored and treated with contempt. Thus, a contextual understanding of skin color and the sociocultural meaning it conveys beyond the physical color properties is required.

Yellow represents the people of Asia; although it does not necessarily represent the physical skin color, it carries a negative connotation ascribed by the Western perspective. The Western society has reduced Asians to a mere yellow color; this is based on the yellow peril, a racist color metaphor that deems the latter as a danger to the former [50][44]. According to Keevak, the Europeans characterized the Chinese and Japanese people as White during their initial contact in the 16th century because they believed that the Asians had the capacity to become Christianized [28]. Toward the end of the 18th century, the term yellow emerged to describe the Asians; moreover, it implied a troubling and threatening connotation when the West and East Asian countries failed to form amicable relations. While "white" was a marker of cultural superiority, the yellow label was associated with inferiority, abnormality, and danger.

The stigmatization of skin color in Asians has persisted thus far. Furthermore, there are noticeable arguments and phenomenon against it as well. Park and Hong critically examined the supposed "whitewashing vs. yellowashing" debates among the international Korean pop (K-pop) fans [35]. Based on the Mibaek esthetics, the majority of the K-pop idols' pictures released by the Korean fans were edited to make their skin appear bright and flawless; nevertheless, some international supporters accused these visuals as "whitewashing" and recolored them to reveal the "natural" skin tones. However, other fans, including those from Korea, alleged these foreign ones of using the "love yourself" rhetoric and circulating photos with yellow-tinted lighting to portray the Koreans' supposed natural skin tone. The debate implied that the matter of Asian complexion has constantly been intertwined with both Orientalism and Occidentalism.

In Japan, where the skin-whitening cosmetics industry was established very early on, the skin-whitening culture has been discussed with respect to its sociohistorical context, rather than an admiration for the West. For instance, Ashikari associated the skin-whitening practice with a modern Japanese gender ideology in which the Japanese middle-class women wear facial makeup publicly [2]. Furthermore, Ashikari also demonstrated it as a way of cultivating the Japanese identity, as tanned skin has been allegedly linked to social minorities such as the Okinawan and Korean individuals [3]. Thus, a light complexion is valued not only as an element of beauty, but also in defining Japaneseness.

Moreover, the belief that "all Asians look the same," which is conventional perception in the West, is currently being challenged by many studies. Wang et al. collected photos of Koreans, Chinese, and Japanese individuals from Twitter to ascertain how they were distinguished from each other using deep learning [55]. In addition, Dobke et al. demonstrated significant differences in the preferred beauty features of the Korean and Japanese women; thus, they argued that diverse ethnic concepts of beauty should be considered, rather than holding an Occidentalized view of Asian beauty [15]. Therefore, the discussion on the differences among all Asians has been receiving attention, as opposed to the tendency to group them together as a certain complexion or appearance.

This study challenged the long-standing notion of the label yellow that has reduced Asians into a race with a homogeneous skin color and otherized them as a minority. By exploring how the East Asians desire to express themselves, we aimed to identify the beauty ideal pursued within their community against the Westernized gaze. Hence, we investigated the East Asians' selfies.

## 2.2 Stereotypical Assumptions in Measuring Facial Colors

Skin colors have been measured in a precisely controlled system. Due to its medical use, lightness, angle, and temperature have been strictly adjusted in laboratory examinations. While the importance of field measurement has arisen, the diversity of participants' composition has encountered certain limitations. Furthermore, the balance of their race has also been considered; however, many previous studies on skin measurements have been conducted in the Western cultures. For example, Hall mentioned that the significance of darker skin experience has been unexplored in social science due to a Eurocentric bias in academic fields [22]. Various skin base colors have been insufficiently examined for the non-Westerners, especially regarding cosmetic use.

For an improved understanding of the hegemonic complexions in the existing color theory, we explored skin color systems such as Pantone and L'Oréal that have been widely employed in the field. Santos explored the Pantone Universe that has dominated the color industry [42]. According to her, Pantone's formation has reinforced the phenomenon of perceiving objects and even human identities through its codes. As mentioned in Santos' research, Dass's work "Humanæ" has exhibited individuals' facial colors using the diversity of the Pantone codes, instead of stereotyped labels such as white, black, and yellow [12].

However, there have been criticisms about the Pantone color system's authority, mainly regarding its monopoly and the propaganda around its annual presenting color, the "color of the year" [30]. To the best of our knowledge, insufficient studies have investigated the density of the color composition of color chips. Not all visible colors have been possibly coded into the Pantone chips; furthermore, we also noticed that the universally standardized color palettes still have limitations in reflecting the human complexions, especially regarding assigning skin color arrays. Practitioners in cosmetic fields have revealed skepticism about using raw Pantone color codes; this study did not measure this specifically as doing so was beyond its scope. According to them, the Pantone's system was unable to sufficiently fill the gap regarding colors desired in the real world, thus requiring emotional aspects to be emphasized. This study performed a detailed comparison between the distributions of hegemonic color systems and the East Asian skin color.

## 2.3 Considering the Social Display Spaces for Measuring Facial Colors

A social media is a source of open personal data. Previous studies have deployed Goffman's self-presentation theory that considers how an individual performs and behaves in social interactions [19][57][7]. The theory uses a framework consisting of two main factors: the player's identity and the personality devised by the play. It has been found to analyze and understand user behaviors and identities as human users curate their virtual identities in social spaces.

As a self-display space, personal photographs on the social medias reveal multiple identities. Xi et al. studied politicians' photos uploaded on Facebook and showed a specific rhetoric [59]. With machine learning training, researchers found that photos could be classified according to the political ideology of individual politicians. We also explored self-images in the social media that contained identities performed by users and which reflected their desires, even if those desires were concealed intentionally. In this respect, social media presents a discursive site where we can discuss the new visibility of the Asians [11]. With the social media technologies enmeshed with self-imaging, Asians display their desired look while refuting the yellow metaphor.

Furthermore, personal photographs from smart devices have recently been applied to skin color measurements. In general, as mentioned above, the facial color for cosmetology or medical uses has been examined under strictly controlled circumstances of light, angle, and temperature. De Regal et al. captured photographs of thousands of women in studio situated in multiple nations [13]. Similarly, studies have recruited people to examine skin colors offline, which was costly and limited by the number of participants [54][9]. Alternatively, Harville et al. used photos to classify skin tones by calibrating the complexion in the photograph without any distortion [23]. Additionally, tracking daily facial colors using smartphones have also been employed for monitoring hemodynamics [38][24]. However, these studies also focused on accurately detecting facial colors, requiring complicated processes and complex models.

This study focused on individuals' intentional facial expressions in their selfies. Therefore, we simply constructed a framework for measuring facial colors and deliberately excluded precise measurements.

## 3 Data & Preliminary Studies

We collected East Asian individuals' selfies that were uploaded on Instagram. By categorizing each image into countries and detecting the skin color, we calculated the tone differences.

First, we focused on the hashtags to find appropriate images on Instagram. We collected the tags based on three criteria: having less noise (such as spam), used properly in each country, and corresponding to selfie images (Table 1). However, regarding the Chinese photos, since Instagram is not officially used in the Chinese Mainland, we attempted to collect those that appeared in Hong Kong. Some languages were encoded into phonetic alphabets owing to technical issues. We gathered images from early 2010 to January 2020, before the wearing of facial masks became compulsory due to the coronavirus pandemic.

Table 1. Hashtags for collecting the data for each country and the number of data.

|  | Query | Number of Data |
|---|---|---|
| South Korean | #얼빡샷 (snap shot) | 3538 |
| Chinese (Based in Hong Kong) | #hkig (Hong Kong Instagram) | 1082 |
| Thailander | #cantik (beautiful) | 7855 |
| Japanese | #自撮り界隈 (my selfie) | 5670 |

Based on the dataset, we first investigated if the selfies could be categorized appropriately, using a deep learning classifier similar to that applied by Wang et al. [55]. We employed our labeled dataset to fine-tune the neural network architecture and examine the extent to which each country could be classified. The network architecture included three convolutional layers, three pooling layers, one dropout layer, and four fully connected layers, resulting in a total of 11 layers. Each convolutional and fully-connected layer had a rectified linear activation function, except for the final one, where a softmax activation function was employed. We applied labeled images for the input data and balanced each label by performing data augmentation.

The neural network-based classification showed a similar performance to the reference model. In brief, the performance of the neural nets was reported to have 91.04% accuracy (loss: 0.2736) for the selfie dataset classification. Specifically, the selfies of each country revealed patterned differences. This classification model aimed to observe whether there were any differences among the countries' selfies in machine perception, neither for building the most precise classification model, nor for making explainable architecture for deep learning outputs. While the deep learning process could not explain the discrepancies in detail, we applied the method to confirm the data features, as performed in reference model.

Our research concentrated on color comparisons. Thus, based on the postcolonial perspective stated above, we focused primarily on skin representation as a component of augmented beauty in selfies.

## 4 Method

In this section, we introduce our framework, which was built for comparative studies on the postcolonial approach to selfies using color analysis. Due to the skin representation defined by tone, texture, and brightness, as Park and Hong discussed, we calculated those features using cheek colors as a source [35]. We built our framework including a facial detection model, a color retrieval model, and clustering techniques, to encode beauty features into numerical values.

For the facial detection, once a face was properly identified in an image, 65 dots were used to detect facial attributes such as eyes, nose, and lips [43]. Since this study aimed to collect a fair complexion, the cheek area was mainly observed. We collected colors from each of the 10 points of the left eyelid edge to the left nasal wing, and the left eyelid's right side to the lips' left edge. Python (version 3.8.5) was utilized for programming and the dlib library (version 19.22.0) for detecting 65 facial attributes.

Not like previous Cutaneous colorimetry researches, we used **hue, saturation, and lightness (HSL) values** to be obtained from the image [26][13]. Both RGB and CIELa*b* used in general, require a complicated conversion process to examine the texture and brightness easily. By analyzing the digital image itself, we were able to create a familiar and straightforward HSL index.

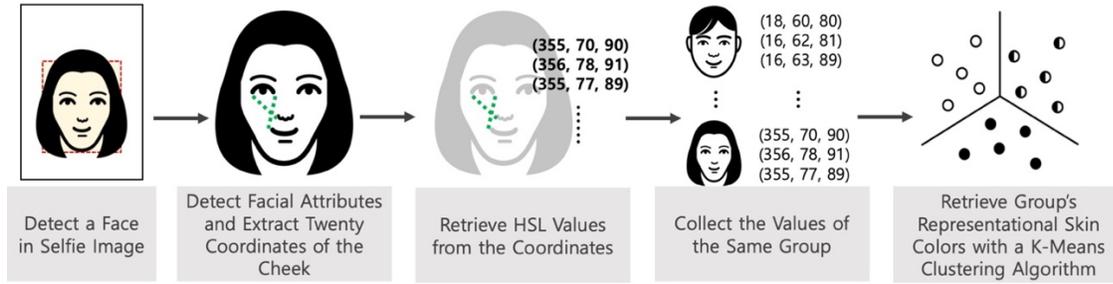

**Figure 1. Process of extracting skin tones from selfies.**

To extract **the representative skin tones** from each country, we employed a k-means clustering algorithm (Figure 1), which, according to Jahanian et al., is commonly used to figure out the contents' color composition in color semantics studies [26]. We created a color palette for each country that contained 20 primary colors agglomerated into each cluster. We only included those colors that had a lightness over 60 (0 (black) to 100 (white)) and a hue between 0 to 50 and 300 to 360, indicating the red areas. We observed that the lightness value under 60 indicated shadows or distorted filters, such as a black-and-white photo conversion. This should be reconsidered in future studies, especially in expanding the number of ethnic groups.

To scrutinize the skin expression, we computed the variance of the HSL values for quantifying the texture. According to Kwon, Korean makeup trends, known as K-beauty, represent flawless porcelain-like skin textures [29]. Therefore, we assumed that a less color variance on faces would indicate fair skin. As shown in formula, we determined the mean of the HSL variances of the photographs.

$$\Delta = \sqrt{(Hue_{(i+1)} - Hue_{(i)})^2 + (Saturation_{(i+1)} - Saturation_i)^2 + (Lightness_{(i+1)} - Lightness_i)^2}$$

**The brightness** of the facial appearance was also measured. We calculated the proportion of bright spots on the face. OpenCV (version 4.5.1), a Python library, recognized the facial zone in the image, and converted the latter into a gray tone. This library picked up considerably bright areas that were nearly white (over 200, 200, 200 based on the RGB color space). The proportion of the bright areas per face indicated how brightly the facial makeup was expressed.

## 5 Findings

In this study, we found that three representations of selfies, namely, skin tone, texture, and brightness, differed among the East Asian countries.

### 5.1 Desired Skin Tone Variations

We observed that the major color compositions varied considerably for each country. Figures 2-1 and 2-2 display the main skin colors extracted using the k-means algorithm in each country group. We assessed the proportion of clusters to visualize the size of each cluster, indicating that those figures could demonstrate the color mainly preferred by users. Figure 2-1, sorted by the size of each color cluster, presents different tones that were primarily preferred in each country. For example, the Chinese's most preferred color, placed in the first column from the left, was more saturated than that of the other countries. Furthermore, the third color from the left in the Japanese palette was brighter than the other colors. We identified slight differences in the colors used mainly among those countries.

As shown in Figure 2-2, in which colors were sorted by lightness, the use of apricot color produced differences among the countries. For instance, the Korean group indicated a smaller proportion regarding the use of light apricot colors (the second column from the left; the HSL value = 18, 0.95, 0.88; proportion as 2.81%) in their selfies, as compared with other countries (for similar color values, the Japanese (2nd column), Chinese (4th column), Thailanders (2nd column) reported 3.27%, 5.47%, and 6.77%, respectively). In addition, pinkish colors, close to the violet ones with the hue values between 300 to 360, were more frequently observed in the Japanese group. They were labeled as "cool tones" in colorimetry, while the red colors, close to yellow, were called "warm tones." Accordingly, the Japanese selfies demonstrated greater appearances with cool tones.

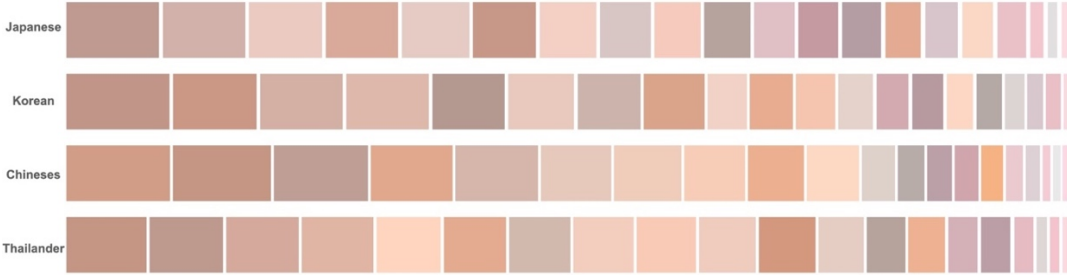

**Figure 2-1. Color palettes primarily extracted based on each country (sorted by the size of each color cluster)**

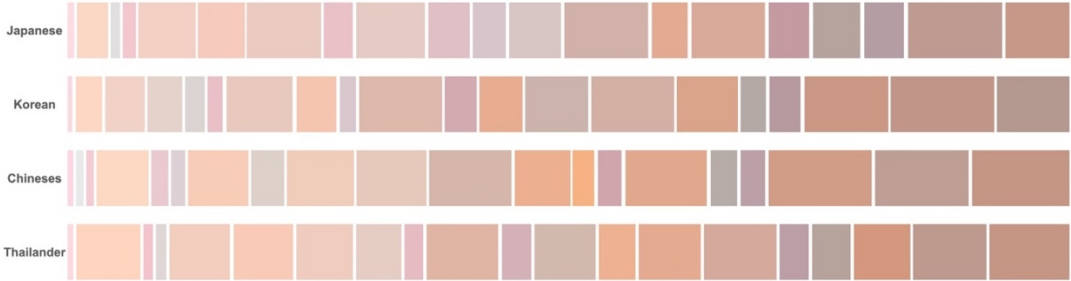

**Figure 2-2. Color palettes primarily extracted based on each country (organized by lightness)**

In the expression of the cheeks' pink color, the Korean and Japanese groups revealed less saturated and smoother shades than the other groups, and their average of saturation values were 0.402 and 0.439, respectively; however, that of the Chinese and the Thailander group were 0.458 and 0.511, respectively. It was inferred that the "Pposhasi" makeup, which uses the Gaussian Blurs and the portraits' opacity adjustment to increase the gray level to decrease the saturation, was actively employed in those countries [36][53]. The Thailander and Chinese groups revealed darker colors of red that were close to orange. In addition to the undertones that appeared to differ among the countries, our findings contradicted the stereotype that "all Asians look the same."

The Euclidean distance revealed the Japanese group to be comparatively distant from the others (0.5651–0.5799). Nevertheless, the Chinese and Korean groups were mostly situated closely to each other (0.5512), indicating that their skin tone preference was relatively similar.

## 5.2  Discrepancy Between the Desired Facial Colors and the Universal Skin Color Themes

In this study, differences were found in the color distributions between the East Asian selfie data and the universally used color systems – Pantone's skin tone guide and L'Oréal's color chart. We inserted the points of all 20 colors extracted from the k-clustering in the country groups in 2D plot, which indicated hue and lightness on the X- and Y-axes. We also marked each group's color scope in existing HSL color wheel on the right, designates hue, saturation, and lightness on the X-, Y-, and Z-axes.

Our results, reflecting East Asia's desires, embraced a wider range of colors than the universally used color systems. As shown in Figure 3, the selfies (greenish dots) showed greater spots and wider extensions to the lighter and cool tones, respectively, which were excluded by Pantone and L'Oréal (reddish dots). The East Asian selfie data revealed the need for a cool toned and brighter face. Moreover, overlapped color ranges also differed among the countries. Regarding the hue range, the Japanese group was placed remotely from the existing systems. This result was consistent with the aforementioned findings that the Japanese selfies demonstrate a cooler tone than the others.

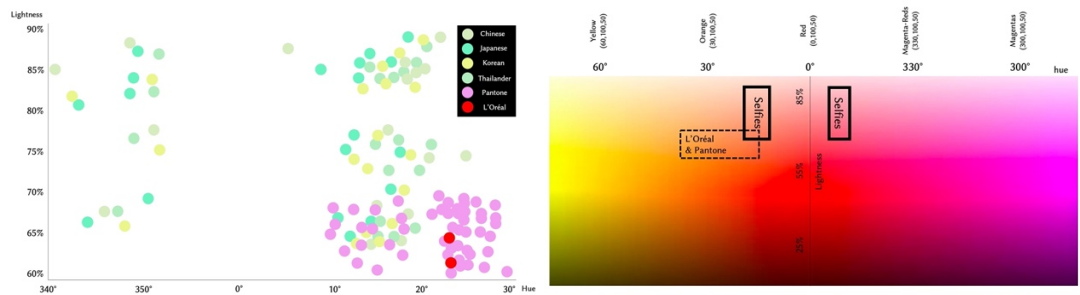

**Figure 3. The East Asian selfies' primary color distribution (greenish dots) and the skin colors of Pantone and L'Oréal (reddish dots). The visualization on the right shows each group's color distribution in HSL color.**

## 5.3  Different Representations of Texture and Brightness

In this study, we noticed that the texture and brightness repertoires in the selfies also varied among the countries. To measure the former, we analyzed the variance in the cheek colors. It was higher in the Japanese group ($p = 0.6846$), as compared to the Korean, Chinese, and Thailander groups (0.5616, 0.5558, and 0.5330, respectively). The flawless skin appearance was more emphasized in the other groups than in the Japanese one.

Notably, the brightness was found to be relatively different from our stereotype. Although K-beauty emphasizes on healthy and porcelain-like facial skin, the proportion of the brightness was indicated to be less in the Koreans (0.002), while the Japanese (0.003), Chinese (0.005), and Thailanders (0.005) employed greater bright spots on their faces. To discover why, we observed each data point to examine the tendency of both texture and brightness expressions.

The observation was inaccurately scrutinized; however, for an enhanced understanding, we have stated our results. It seemed that the Japanese people used a higher contrast of facial blusher on their cheeks and did not conceal dots or freckles on their faces. Furthermore, they actively utilized augmented stickers that covered their cheeks on the selfie. Contrariwise, while the Thailanders and the Chinese did not use colorful blushers actively, they positively employed lights while capturing portraits. In future studies, this method should be refined to obtain detailed results.

Thus, we expect that our method can infer the makeup style or behavior from selfies. If the stickers, hands, masks, and hair could be clearly removed from photos, the tendency of skin expression with respect to brightness and texture could be researched quantitatively.

# 6  Discussion: Selfies, Desires, and the Augmented Skin Color

This study found that facial skin color desires differed among the East Asian countries. The tone showed a discrepancy among them, especially the Japanese pattern in the hue channel. Also our results demonstrated that selfies could act as a source for desire analysis. While the selfie itself represents the augmented production of the user's beauty preference, our methodological approach could be applied to cosmetic trend analyses. Owing to technical advances in computer vision, we can utilize facial recognition and automatic color extraction with considerably greater ease than done previously. We suggest this method at the work site. The following section discusses the implications of our findings.

## 6.1 Different Desires and New Imaginations

There was a case where one global cosmetic brand mall sent ivory nude color powder instead of the customer's demand for a shell pink one. This decision was based on the market's intuitive thinking regarding the "supposedly suitable" color for the Asians; however, it faced the criticism of stereotypical racism with respect to Asians' yellow faces. The postcolonial implications embedded in our findings could be helpful for overcoming this kind of deep-entrenched racial stereotypes in beauty market.

As shown in Figure 1, the desired skin expression in the virtual world and the real complexion in the mirror vary. Because the augmented skin appearances in social media have an emotional aspect, this approach could overcome the stereotyped and homogenized imagination of Asian looks. Previous research asserted that "East Asians look different"; our finding, that "East Asians' desires vary," supplemented their claims. It is associated with the Asian appearances that the Western media exhibit. The makeup style of an extreme shading of the cheek contour and emphasizing the slant eyes, such as that animated by Disney in Mulan (1998), evokes antipathy in Asians. Even in Asian countries, they use different makeup styles. The Koreans highlight down-toned natural makeup to their dewy faces, while the Japanese emphasize the "Gyaru" makeup, which focuses on utilizing color makeup on matte skin.

The existing skin color system in global beauty industry lacks tool nor sensitivity to represent different desires of Asians. Meanwhile, augmented skin color created by social media and selfie technologies are conveying the desires of Asians. Their desires do not accord with Western imagination, and, more significantly, desires are always plural.

## 6.2 Emotional Perspective to the Customized Beauty Systems

Hence, this study contributed to the methodological suggestion having a multidimensional approach to the beauty system. Beyond the tone-based analysis of skin appearances, brightness and texture were also significant to the study. They reflected the changeable flow in makeup trends in which the glowing and dewy skin are weighed. The perception of the beauty system's other components would contribute to overcoming the limitations of a cosmetic trend analysis. By employing quick and simple technical aids, marketers can flexibly cope with the rapidly changing trends.

Recently, the cosmetic market has faced criticisms for being biased, as described above. Thus, delicate customizations have been widely applied in the field. In designing a personalized beauty system, we expect that our method will bring attention to users' cultural backgrounds. An existing service has recommended a personal skin foundation based on a user's selfie using an undertone extraction with a white-balancing algorithm [32]. We suggest an emotional perspective by considering the user's cultural and communal preferences.

The emotional perspective can be differentiated from the view of previous systems as we discussed above. While selfies represent personal beauty preferences, using personally edited portraits could be dealt as a mirrored data for understanding current emotional mood. We, thus, found possibilities to measure emotions through colors in selfies in this research. By matching with color psychology, our method can be further utilized in the sequential emotion analysis of individual users and the society. We could examine the variance of mood preferences of the individual user and the entire community. With the flow analysis, the psychological and pathological states could be predicted. Moving further from the physical aspect, we integrated the inner state analysis into our approach.

Furthermore, social trends and societal effects could also be measured by sequentially analyzing an individual's desired color tracking. By conducting mixed studies with media content or market data, we could determine the origin of the individual's desire.

## 6.3 Design Considerations for the Selfie Filter Apps

Based on these data, we should consider the basis of a selfie. We strongly suggest that diversity should be applied to the default modes of beauty filters and smartphone cameras. Most apps are developed based on the users' beauty preferences; however, the reverse effect caused by those programs cannot be disregarded. As debated by Gomez et al., developers should consider filter-drop options for their camera functions [20].

Because these programs cover the societal values, the hazardous side of big data, especially in terms of biases, should be questioned seriously. In the big data era, a hegemony can be built on machine decisions affected by data composition.

To prevent data imbalance, the ethnic group, gender, and age should be considered in the face attribute dataset [8][27]. As a recommendation to the beauty system, we suggest that the big data employed in the algorithm should be additionally specific and deeply collected at the geographical or cultural levels to avoid suspicions such as stereotypes toward Asians.

Considering that the literacy of users is not the same, manufacturers and publishers of phones and apps should pay attention to the beautifying system's criteria reflected in their products. However, Rhue asserted that the output of the algorithm could have an anchoring effect on human decisions [40]. Moreover, Steed and Caliskan urged that the algorithm pre-trained with human biases could enforce human stereotypes [48]. Therefore, beauty filter apps should clearly announce their architectural state to the public. Whether a hegemonic system emerges from our daily use products structured on a stereotyped basis should be monitored.

# 7 Conclusions

Using a selfie analysis, this study found that the skin representation among East Asians varies across countries. Its findings can be summarized as follows: Despite being classified as a uniform group, the East Asians showed considerably different values regarding the tones, textures, and brightness of facial photographs in each country. Specially, the Japanese group demonstrated considerably greater distances from the Korean, Chinese, and Thailander groups in the texture analysis. Furthermore, the East Asian's selfie colors did not correspond with the existing skin color system. Higher lightness values and a broader range of hues were observed in the present study.

Based on these findings, this study makes three contributions. First, we suggested a method to determine preferences and desires for beauty by using social media data. This technique is considerably simpler than the existing skin measurement methods, and is more appropriate for verifying trends quickly. Second, we provided a desired skin representation system for each East Asian country that differs from and overcomes the Westernized yellow stereotype. This research resolved the discordance between the East Asian customers' desires and the existing cosmetic system. Finally, our visualization of the out-of-range color values to the dominant skin color system could provide clues for beauty preference analyses, which were unobserved in the actual skin measurements.

This study has certain limitations. First, the texture and brightness analyses should be more delicately designed in the in-the-wild dataset. We expect this method to be more useful with additionally precise noise-off algorithms. Second, the threshold for skin tone was exceedingly subjectively determined. We referred to the East Asian skin color as over 60% light; however, our decisions could be racially biased. Finally, the range of "seem optically identical colors," as suggested by MacAdam, should be reconsidered with our findings [33]. Colors that are physically and relatively close may be perceived as almost the same color. An alternative should be developed to effectively demonstrate our results. In the protection of personal data, the effect of our method could also be limited. We should consider how to derive user acceptance to make themselves beautify with greater satisfaction.

This study primarily aimed to determine the different values of individuals regarding skin appearances. In future research, we aim to expand our study in three layers. First, it should investigate how the growing makeup trends, such as K-makeup, indicate different patterns in each continent. We predict that meaningful results could be obtained by analyzing the relationship between the media and the rising patterns of beauty desires. Second, the research can be extended to in-field studies in cosmetic markets, including production lines and makeup artists. On the market's side, who need to lead the trend, a more scientific and data-driven methodology should be considered. Our approach can be developed more reliably in the field. Finally, in the near future, our findings in this study should be further discussed regarding the hegemonic aspect from the perspective of Dyer's Whiteness. The study could be conducted cross-culturally and by analyzing the difference from "seen-as-white." Contrary to using macro-lenses in this study, future research should deploy a wide-angled view to determine the augmented desires of a various range of individuals.

We continue to investigate how East Asians chase the emotional goal based on the discrepancies between their actual skin and preferred colors. By revealing their augmented desires, the unfamiliarity in the virtual identity and reality could be filled in.

## Declaration of Conflicting Interests
The author(s) declared no potential conflicts of interest with respect to the research, authorship, and/or publication of this article.

## Ethical Statements
This manuscript has not been published or presented elsewhere in part or in entirety and is not under consideration by another journal. All study participants provided informed consent, and the study design was approved by the appropriate ethics review board. We have read and understood journal's policies, and we believe that neither the manuscript nor the study violates any of these.

## Acknowledgements
The authors thank the anonymous reviewers for their constructive feedback.

[38] Louisa F. Polania, Raja Bala, Ankur Purwar, Paul Matts, and Martin Maltz. 2020. Skin Chromophore Estimation from Mobile Selfie Images using Constrained Independent Component Analysis. IS&T International Symposium on Electronic Imaging, HTTPS://DOI.ORG/10.2352/ISSN.2470-1173/2-2-/14/COIMG357

[39] Rankin. 2019. Selfie-Harm Project. Retrieved September 15, 2022 from http://www.visualdiet.co.uk/selfie-harm/

[40] Lauren Rhue. 2019. Beauty is in the AI of the Beholder: How Artificial Intelligence Anchors Human Decisions on Subjective vs. Objective Measures. Fortieth International Conference on Information Systems, Munich 2019.

[41] Francesca C. Ryding and Daria J. Kuss. 2020. The Use of Social Networking Sites, Body Image Dissatisfaction, and Body Dysmorphic Disorder: A Systematic Review of Psychological Research. Psychology of Popular Media, Vol. 9, no. 4, 412-435. HTTPS://DOI.ORG/10.1037/ppm0000264

[42] Roberta Schultz Santos. 2014. PANTONE: Identity Formation Through Colours, Thesis for Master fo Arts, Contmeporary art, design and new media art histories in OCAD University, Toronto, Ontario, Canada, April 2014.

[43] Christos Sagonas, Epameinondas Antonakos, Georgios Tzimiropoulos, Stefanos Zafeiriou, and Maja Pantic. 2016. 300 Faces In-The-Wild Challenge: database and results. Image and Vision Computing 47, 3-18. HTTPS://DOI.ORG/10.1016/j.imavis.2016.01.002

[44] Edward W. Said. 1978. Orientalism. Routeledge.

[45] Sara Santarossa and Sarah J. Woodruff. 2017. #SocialMedia: Exploring the relationship of social networking sites on body image, self-esteem, and eating disorders. Social Media + Society, April-June 2017, 1-10. HTTPS://DOI.ORG/10.1177/2056305117704407

[46] Junho Song, Kyungsik Han, Dongwon Lee, and Sang-Wook Kim. 2018. "Is a picture really worth a thousand words?": A case study on classifying user attribtes on Instagram. PLoS ONE 13(10):e0204938. https://doi.org/10.1371/journal.pone.0204938

[47] Gayatri Chakravorty Spivak. 2003. Can the subaltern speak?. *Die Philosophin*, *14*(27), 42-58.

[48] Ryan Steed and Aylin Caliskan. 2021. Image Representations learned with unsupervised pre-training contain human-like biases. FaccT'21. March 3-10, 2021. Virtual Event. HTTPS://DOI.ORG/10.1145/3442188.3445932

[49] The Korea Times. Estee Lauder accused of racism in product switch for Asian customers. (2020.11.19) Retrieved April 6, 2021 from http://www.koreatimes.co.kr/www/art/2020/11/199_299011.html

[50] Richard Austin Thompson. 1978. The yellow peril, 1890-1924. New York: Arno Press.

[51] Jolanda Veldhuis, Jessica M. Alleva, Anna J. D. Bij de Vaate, Micha Keijer, and Elly A. Konjin. 2020. Me, m selfie, and I: The relations between selfie behaviors, body image, self-objectification, and self-esteem in young women. Psychology of Popular Media, 9(1), 3-13. HTTPS://DOI.ORG/10.1037/ppm0000206

[52] Liwei Wang, Yan Zhang, and Jufu Feng. 2005. On the Euclidean Distance of Images. IEEE Transactions on Pattern Analysis and Machine Intelligence, Vol. 27, No. 8.

[53] Liping Wang, Chengyou Wang, and Xiao Zhou. 2017. Blind Image Quality Assessment on Gaussian Blur Images. Journal of Information Processing Systems, Vol. 13, No. 3, pp.448-463. June 2017. HTTPS://DOI.ORG/10.3745/JIPS.02.0059

[54] Mengmeng Wang, Kalda Xiao, Sophie M Wuerger, Vien Cheung, and Ming Ronnier Luo. 2015. Measuring Human Skin Colour. Proceedings of the 23rd Color and Imaging Conference.

[55] Yu Wang, Haofu Liao, Yang Feng, Xiangyang Xu, and Jiebo Luo. 2016. Do they all look the same? Deciphering Chinese, Japanese and Koreans by fine-grained deep learning. https://doi.org/10.48550/arXiv.1610.01854

[56] Katie Warfield. 2014. Making Selfies / Making Self: Digital Subjectivities in the Selfie. On-site presentation at the fifth international conference on the image and the image knowledge community, freie Universität, berlin, germany. Oct 29-30, 2014.

[57] D.E. Wittkower. 2014. Facebook and Dramauthentic Identity: A Post-Goffmanian Model of Identity Performance on SNS. First Monday 4-7. HTTPS://DOI.ORG/10.4210/fm.v19i4.4858

[58] Christine T. Wolf, Haiyi Zhu, Julia Bullard, Min Kyung Lee, and Jed R. Brubaker. 2018. The Changing Contours of "Participation" in Data-driven, Algorithmic Ecosystems: Challenges, Tactics, and an Agenda. Companion of the 2018 ACM Conference on Computer Supported Cooperative Work and Social Computing. 377-384.

[59] Nan Xi, Di Ma, Marcus Liou, Zachary C. Steinert-Threlkeld, Jason Anastasopoulos, and Jungseock Joo. 2019. Understanding the political ideology of legislators from social media images. Proceedings of the International AAAI Conference on Web and Social Media, vol 14. 726-737.